\documentclass[%
 reprint,
superscriptaddress,
 amsmath,amssymb,
 aps,
]{revtex4-1}

\newcommand{\br}{\mathbf{r}}

\newcommand{\CX}{\mathcal{X}}

\newcommand{\CA}{\mathcal{A}}

\usepackage{graphicx}%
\usepackage{dcolumn}%
\usepackage{bm}%
\usepackage{verbatim}
\usepackage[version=4]{mhchem}
\usepackage{xcolor}
\usepackage{physics}
\usepackage[normalem]{ulem}
\usepackage{pdfpages}
\usepackage{pgffor}

\makeatletter
\AtBeginDocument{\let\LS@rot\@undefined}
\makeatother

\begin{document}

\preprint{APS/123-QED}

\title{
A New Kind of Atlas of Zeolite Building Blocks
}%

\author{Benjamin A.\ Helfrecht}
 \affiliation{Laboratory of Computational Science and Modeling, Institut des Mat\'eriaux, \'Ecole Polytechnique F\'ed\'erale de Lausanne, 1015 Lausanne, Switzerland}

\author{Rocio Semino}
\affiliation{Institut Charles Gerhardt Montpellier UMR 5253 CNRS, Universit\'e de Montpellier, Place E. Bataillon, 34095 Montpellier Cedex 05, France}
 \affiliation{Laboratory of Computational Science and Modeling, Institut des Mat\'eriaux, \'Ecole Polytechnique F\'ed\'erale de Lausanne, 1015 Lausanne, Switzerland}
\author{Giovanni Pireddu}
\affiliation{Dipartimento di Chimica e Farmacia, Universit\`a degli Studi di Sassari, Via Vienna 2, 01700 Sassari, Italy}
 \affiliation{Laboratory of Computational Science and Modeling, Institut des Mat\'eriaux, \'Ecole Polytechnique F\'ed\'erale de Lausanne, 1015 Lausanne, Switzerland}

\author{Scott M.\ Auerbach}
\affiliation{
 Department of Chemistry and Department of Chemical Engineering, University of Massachusetts Amherst, Amherst, MA 01003 USA\\
}

\author{Michele Ceriotti}
 \affiliation{Laboratory of Computational Science and Modeling, Institut des Mat\'eriaux, \'Ecole Polytechnique F\'ed\'erale de Lausanne, 1015 Lausanne, Switzerland}
\email{michele.ceriotti@epfl.ch}

\date{\today}%

\begin{abstract}
We have analysed structural motifs in the Deem database of hypothetical zeolites, to investigate whether the structural diversity found in this database can be well-represented by classical descriptors such as distances, angles, and ring sizes, or whether a more general representation of atomic structure, furnished by the smooth overlap of atomic positions (SOAP) method, is required to capture accurately structure-property relations. 
We assessed the quality of each descriptor by machine-learning the molar energy and volume for each hypothetical framework in the dataset. We have found that SOAP with a cutoff-length of 6 \AA{}, which goes beyond near-neighbor tetrahedra, best describes the structural diversity in the Deem database by capturing relevant inter-atomic correlations. Kernel principal component analysis shows that SOAP maintains its superior performance even when reducing its dimensionality to those of the classical descriptors, and that the first three kernel principal components capture the main variability in the data set, allowing a 3D point cloud visualization of local environments in the Deem database. This ``cloud atlas" of local environments was found to show good correlations with the contribution of a given motif to the density and stability of its parent framework. Local volume and energy maps constructed from the SOAP/machine-learning analyses provide new images of zeolites that reveal smooth variations of local volumes and energies across a given framework, and correlations between local volume and energy in a given framework.
\end{abstract}

\maketitle

\section{\label{sec:introduction} Introduction}

Zeolites are nanoporous, crystalline silica-based materials featuring a rich array of structures \cite{ZeoHandbook2003}.
They are primarily composed of corner-sharing \ce{SiO4} tetrahedra, and may include heteroatoms such as \ce{Al}, \ce{Ge}, and \ce{P} isomorphically substituted for \ce{Si} sites. Because of their stable frameworks and versatile structures and compositions, zeolites have enjoyed many applications---especially as catalysts in shape-selective transformations of hydrocarbons, and as molecular sieves for separating mixtures.
Databases of real \cite{ZeoAtlas2007} and hypothetical \cite{Deem2009,Pophale2011,Treacy1997,Treacy2004,Li2013} zeolites have been screened for new applications such as carbon dioxide storage \cite{Lin2012}, showing the potential utility of several hypothetical zeolites. 
Despite this promise, synthesizing new, tailor-made zeolite frameworks has remained a grand challenge in materials science, largely because we lack fundamental understanding of how zeolites crystallize in solution. 
To shed light on this, recent work has considered assembly \cite{Lupulescu2014,Kumar2018} of various rings \cite{Zhu2016} and cages \cite{Blatov2013} inspired by known or hypothetical zeolites. 
While such approaches are logical, enumerating zeolite sub-structures through the lens of known rings and cages can leave out many conceivable local silica environments not yet encountered, which may be synthesizable \cite{Li2013} and also important for investigating disordered silica structures leading up to zeolite crystals. In pursuit of a more bias-free understanding of zeolite structures, we implement herein the smooth overlap of atomic positions (SOAP) \cite{bart+13prb,de+16pccp} descriptor to represent local silica environments in the Deem database of hypothetical zeolite frameworks, and critically compare this descriptor to several well-established descriptors of zeolite structure.  

Real and hypothetical zeolite structures are currently understood in terms of framework topologies, which are mathematical connectivities of corner-sharing tetrahedra. These topologies are analyzed, in turn, by identifying rings that form each structure, where an ``$n$-ring'' involves $n$ alternating \ce{-Si-O-} atomic units. At the time of this writing, there are 234 topologies (of which 223 are fully connected) of fully ordered zeolite materials in the online database of the International Zeolite Association \cite{ZeoAtlas2007}, each labeled by a three-letter code. For example, FAU labels a topology with 12-rings exhibited by actual materials such as faujasite, NaX, and NaY, while MFI labels a topology with 10-rings exemplified by zeolites silicalite (all-silica) and ZSM-5. 
In this database there are roughly 40 topologies that can be synthesized as all-silica zeolites, i.e., polymorphs of $\alpha$-quartz. 
Forcefield calculations on many of these all-silica zeolites have found them to be 8.0--20.\ kJ/mol \ce{SiO2} less stable than $\alpha$-quartz, with energies that correlate linearly with framework density \cite{Henson1994}, in good agreement with experimental calorimetry \cite{Petrovic1993}.

The Atlas of Zeolite Structures also contains a collection of 60 zeolite building blocks known as composite building units (CBUs) \cite{ZeoAtlasCBUs}, which are motifs such as the double 4-ring (``d4r'') and the double 6-ring (``d6r''). The overwhelming majority of CBUs are cage-like objects centered on empty space. In the present article, we take a complementary approach to identifying zeolite building blocks by constructing {\emph{atom}}-centered environments rather than void-centered environments. We pursue such atom-centered environments because we find it simpler logically and mathematically to reconstitute zeolite frameworks and their properties by summing over atoms instead of summing over void spaces. In addition, this construction represents a natural extension of well-established approaches based on fragment decomposition of molecules, and reflects the physical principle that macroscopic properties of materials arise from the combination of local contributions from their atomic or molecular components. 

In addition to the database of known all-silica zeolites, there exist databases of hypothetical all-silica zeolites containing millions of candidate framework structures \cite{Deem2009,Pophale2011,Treacy1997,Treacy2004,Li2013}. For example, the database constructed by Deem and co-workers includes fully connected, all-silica zeolitic structures with energies within 30 kJ/mol \ce{SiO2} of $\alpha$-quartz \cite{Pophale2011}, as computed with the shell-model-based Sanders--Leslie--Catlow (SLC) forcefield \cite{Sanders1984,Schroder1992} in the program GULP \cite{Gale2003}. While these structures share common local structural features such as \ce{Si-O} bond lengths near 1.6 \AA{} and \ce{O-Si-O} angles near 109.5$^\circ$, they also exhibit substantial diversity. This diversity is often characterized by distributions over descriptors such as \ce{Si-O-Si} angles \cite{Zones2011}, \ce{Si-Si} near-neighbor distances, and ring sizes \cite{Deem2009,Earl2006,Pophale2011,Curtis2003}. 
In this article, we address the question of whether such descriptors are optimal or even sufficient for describing the structural diversity found in the Deem SLC-PCOD database (which we refer to as simply the ``Deem database''), by comparing these conventional zeolite descriptors with a more general geometric representation of network structure focused on atom-centered environments.

To perform this comparative assessment of structural descriptors, we require the following two components: (1) an alternative mathematical description of zeolite structure that is concise yet complete, which incorporates all of the trivial physical symmetries (e.g., translational and rotational symmetries and invariance upon permuting atom numbers) and that is sufficiently general to be applicable to all known structures, as-yet-undiscovered structures, and even disordered structures that may lead to zeolites; (2) an objective way of assessing the performance of a given structural representation in mapping the relevant structural diversity~\cite{ceri19jcp}.

The recent surge of applications of machine learning (ML) to chemistry and materials science, that includes accurate predictions of molecular properties\cite{Bartok2017,Ward2017,Faber2015,Faber2016,Faber2017,Hansen2013,Hansen2015,Schutt2014,Rupp2012,vonLilienfeld2015,ramp+17npjcm,Isayev2017,de+16pccp,Gastegger2018}, the construction of data-driven interatomic potentials~\cite{Behler2011,bart+10prl,bart+13prb}, as well as fundamental studies of the mathematical representation of atomic structures,~\cite{glie+18prb,will+19jcp,drau19prb}  provides several suitable mathematical descriptions satisfying criterion (1) above.
Previous applications of ML in zeolite science, which include identifying zeolite framework topologies \cite{Yang2009} and predicting mechanical properties \cite{Evans2017}, have relied on zeolite-specific structural descriptors. Our goal is to establish a more general structural representation, which could be useful for describing both crystal structures and disordered materials leading up to crystals.
In particular, we choose the smooth overlap of atomic positions (SOAP) representation~\cite{bart+13prb}, which has been used successfully to construct interatomic potentials~\cite{Bartok2015,Szlachta2014,Bartok2018,Bartok2010,Maillet2018,Deringer2017,Dragoni2018}, to predict the properties of atomic-scale structures or identify motifs in complex materials such as molecular crystals \cite{Musil2018} and polypeptides \cite{helf+19fmb}. 
SOAP describes local atomic environments in terms of spherically invariant measures of the neighbor distribution around each atom, as described in more detail in the next section, thus providing the atom-centered approach discussed above. In the present article, we investigate the applicability of SOAP for elucidating the structural diversity in the Deem SLC-PCOD database of hypothetical zeolites. 
This choice is motivated by the very large number of hypothetical, all-silica frameworks that are present in this database (as compared to the ~40 frameworks in the IZA database), which allows for ample statistics and thus a proper machine learning study.

To objectively assess the performance of a given structural descriptor, we require 
that it leads to an effective machine learning model to predict key materials properties~\cite{ceri19jcp}.
Below we show that the SOAP-based representation includes more information than standard descriptors such as \ce{Si-O-Si} angles, \ce{Si-Si} near-neighbor distances, and ring sizes as evidenced by more accurate machine learning predictions of zeolite molar volumes and lattice energies, which are fundamental properties of zeolite frameworks \cite{Henson1994}. By comparing the predictions of zeolite volume and energy based on local atomic environments with different nominal sizes, we also discover the degree of locality of each property, i.e., the correlation lengths that are required to determine the overall behavior of that particular property. We also show below new ways to visualize zeolite frameworks in terms of local contributions to energy and volume.

The remainder of this article is organized as follows: in Section~\ref{sec:methods} we provide a concise introduction to the SOAP framework, and to the supervised and unsupervised ML algorithms we adopt; in Section~\ref{sec:data} we describe the dataset we used to demonstrate our construction; in Section~\ref{sec:results} we present and discuss our results, and in Section~\ref{sec:conclusions} we draw our conclusions. 

\section{\label{sec:methods} Methods}
We start by presenting an overview of the methods we use to represent structural environments in silica-based frameworks---which we define in terms of overlapping motifs centered on each \ce{Si} atom in the 
framework---and to investigate quantitatively the relation between such environments and some macroscopic properties of each framework, namely the molar volume and energy. 

\subsection{\label{sec:classical} Classical Descriptors}
In the context of zeolitic structure analysis, the use of geometric descriptors for classifying and rationalizing structure-property relationships is already well established \cite{Smith1963,ZeoAtlas2007,ZeoHandbook2003,LeRoux2010}. The choice of the representation for a given zeolitic structure is often motivated by physico-chemical understanding and intuition regarding which structural features are relevant to the study of certain properties. For example, when investigating catalytic properties of zeolites, one may consider correlating acid-site strengths with \ce{Si-O-Al} angular distributions. On the other hand, when considering diffusion of guest molecules through zeolite frameworks, one typically uses zeolite ring distributions to rationalize transport properties. We refer to such representations as \emph{classical descriptors} to emphasize the difference between zeolite-specific descriptors and general descriptors such as SOAP.

We focus in particular on three widely used classical descriptors to serve as references for the chemical environment representations: \ce{Si-O-Si} angles, \ce{Si-Si} distances, and connected rings.
The distance- and angle-based descriptions are local, atom-centered features, and are appropriate for representing properties that could be safely decomposed into additive, local contributions. For instance, bonding interactions in interatomic potentials are often expressed in terms of bond angles and distances \cite{Stoneham1986,Balamane1992}.
The distance- and angle-based representations characterize the local environment of each \ce{Si} as a vector of the \ce{Si-Si} distances or \ce{Si-O-Si} angles between the central reference \ce{Si} atom and the four nearest \ce{Si} atoms, as these structures are mainly composed of \ce{Si} tetrahedra. 
In order to make these representations independent of permutations of the atom indices, the vector elements are arranged in descending order.

Ring-based descriptors involve correlations on longer length scales, and consequently may be more suitable for characterizing the topology of a given framework than distance- or angle-based structural representations. 
Zeolitic frameworks can be described in terms of rings according to various definitions; the ring  classifications included in the IZA database are those required to build up a given structure, plus those required to define pores and channels accessible to guest molecules. Such ring systems are readily confirmed by visual inspection of the crystal structure. 
To apply an automated analysis to a large database of structures, we use two mathematically rigorous definitions in our work, namely, King's criterion \cite{King1967,LeRoux2010} 
and the shortest path criterion \cite{Guttman1990,Marians1990,Franzblau1991,LeRoux2010}, as implemented in the R.I.N.G.S. code~\cite{LeRoux2010,RINGS}.
In both cases, we translate the list of detected rings in a given zeolite framework into vectors of features $x_s$ associated with local \ce{Si} environments by counting the number of times the tagged \ce{Si} atom appears in $s$-sized rings (since O atoms are also present in the rings, the real ring size is $2s$, here we adopt the frequently used convention of naming a ring counting only \ce{Si} atoms). For example, the ring vector for a \ce{Si} atom in silica-sodalite (which are all equivalent by symmetry) is $[0, 0, 0, 2, 0, 4, 0, \dots, 0]$ indicating that each Si atom is part of two 4-rings and four 6-rings. In general, all ring vectors in the present work are 10-dimensional vectors, ranging from 3-rings to 12-rings. In the present work we focus on the shortest-path definition, which gave marginally better performance than King's when used as the basis for predicting structure--property relations.

\subsection{\label{sec:SOAP} The SOAP representation}
\begin{figure}
    \centering
\includegraphics[width=0.9\columnwidth]{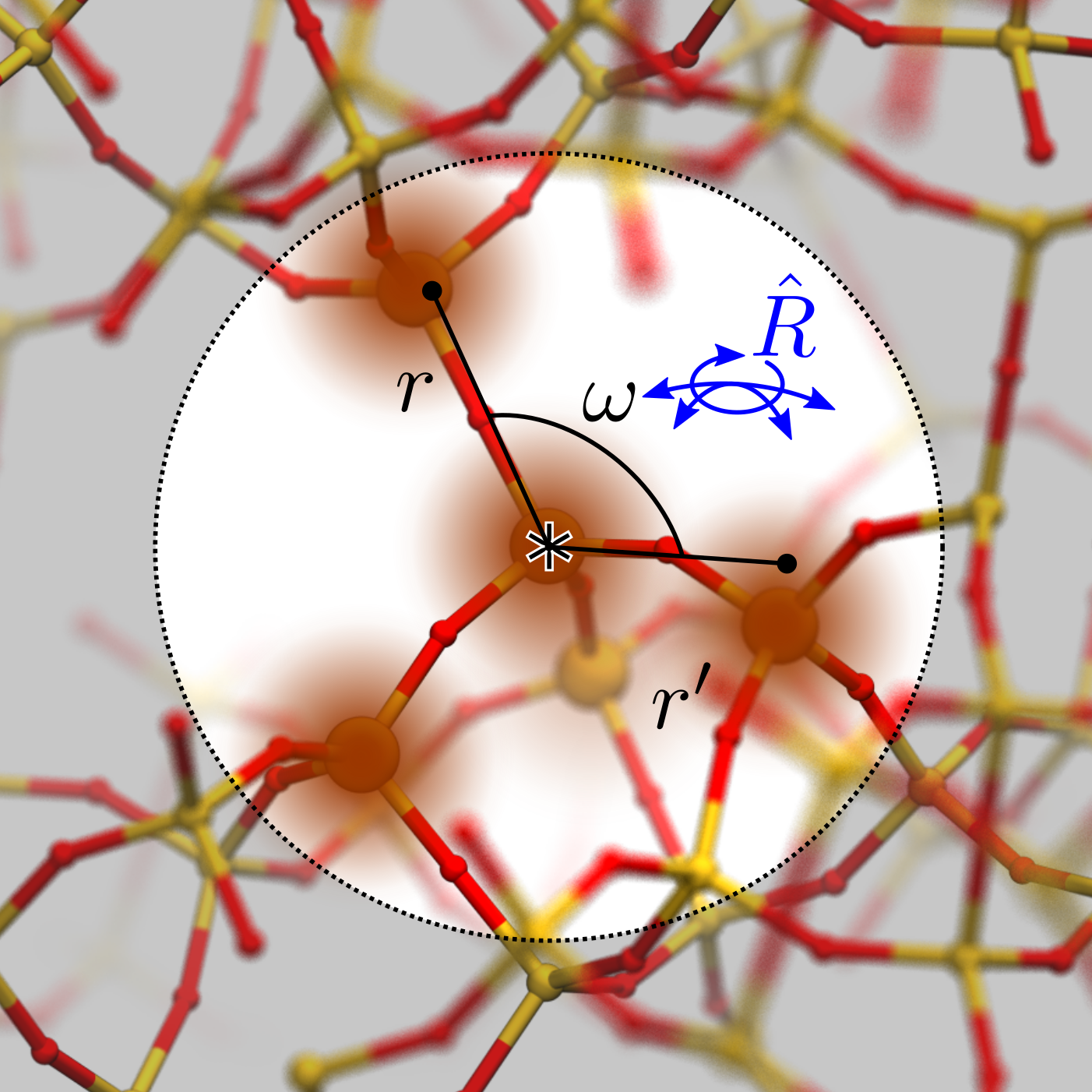}
\caption{Schematic of the SOAP representation applied to a local \ce{Si} environment in the zeolite CDO. A smooth atom density (here represented only for Si atoms) is built as a superposition of atom-centered Gaussians, and a smooth three-body correlation function (as a function of two distances $r$ and $r'$, and of the angle between the two directions $\arccos\omega$) is computed by evaluating this density in two points, and averaging the result over all orientations of the $(r,r',\omega)$ stencil.}
    \label{fig:soap-scheme}
\end{figure}

The SOAP representation can be seen in relation to the symmetrised three-body atom correlation function, defined for an environment $\CX_j$ centered around the $j$-th atom in a structure~\cite{will+19jcp}. 
One can start by building a smooth atom density centered on atom $j$, separated in different channels depending on the chemical species $\alpha$ (\ce{Si},\ce{O} in this case) as follows:
\begin{equation}
\psi^\alpha_{\CX_j}(\br)=\sum_{i\in\CX_j^\alpha} g(\br-\br_{ij}) f_\text{c}(r_{ij}),
\label{SOAP1}
\end{equation}
where Eq.\ (\ref{SOAP1}) involves a sum of Gaussian functions centered on the interatomic distances $\br_{ij}$, which are restricted to atoms of type $\alpha$ in the vicinity of atom $j$ using a smooth cutoff function, $f_\text{c}(r_{ij})$. For the cutoff function, we chose a section of a cosine function switching from 1 to exactly zero within 0.5\AA{} of the cutoff distance.
The (Gaussian-smoothed) three-body correlation function can be defined as an average over all rotations operators $\hat{R}$ of the density evaluated at two different points
\begin{equation}
\begin{split}
\bra{\alpha r \alpha' r'\omega}\ket{\CX_j} = &\int \mathrm{d}\hat{R}\, rr' \psi^\alpha_{\CX_j}(r \hat{R}\hat{\mathbf{e}}_z) \times \\ \times &\psi^{\alpha'}_{\CX_j}(r' \hat{R}(\omega \hat{\mathbf{e}}_z +\sqrt{1-\omega^2}\hat{\mathbf{e}}_x)), 
\end{split}
\end{equation}
which corresponds to the  probability of finding two atoms of type $\alpha$ and $\alpha'$ at distances $r$ and $r'$ from a tagged central atom, in directions forming an angle $\arccos \omega$. In this expression, the density is evaluated at two points, defined in terms of the unit vectors along the $z$ and $x$ axes of a Cartesian reference system.
This construction is schematically depicted in Fig.~\ref{fig:soap-scheme}, which shows the Gaussian atom density built on the neighbors of a central atom, and the two points at which the density is evaluated. For each value of $(r,r',\omega)$ the resulting density correlation is averaged over all the possible orientations of the points in space.

\begin{figure*}
    \centering
\includegraphics[width=1.0\linewidth]{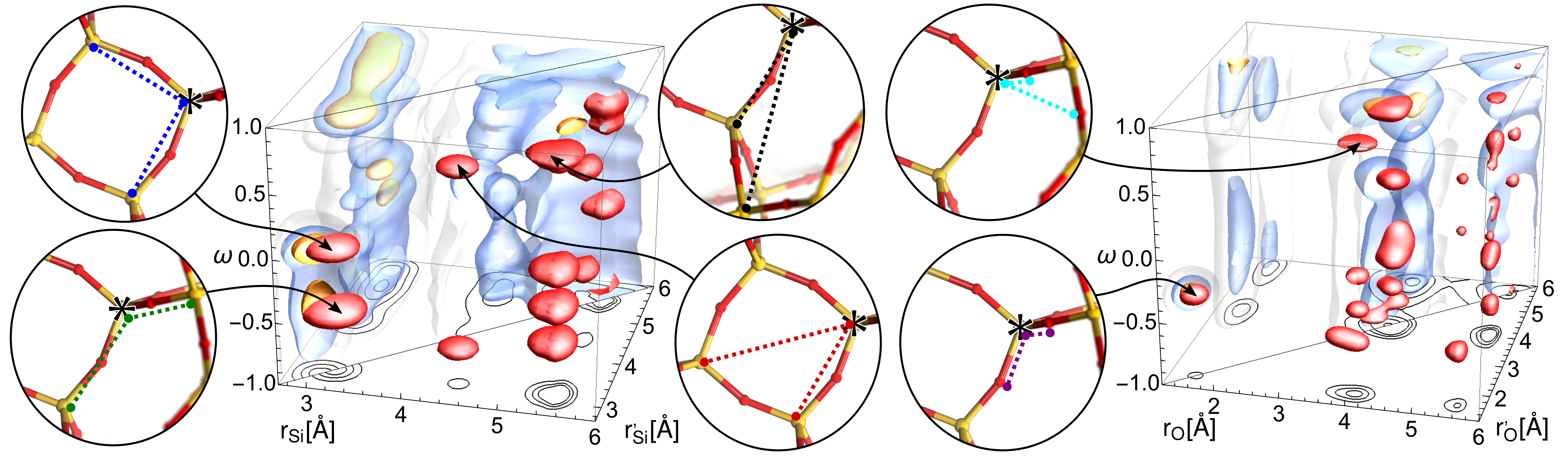}
\caption{Contour plots corresponding to the 3-body correlation function for \ce{Si-Si} (left) and \ce{O-O} (right) atoms around a central silicon atom. White, blue and orange contours (drawn for $r<r'$) correspond to three isocontours of the density averaged over 1,000 structures from the database. Red isocontours  (drawn for $r>r'$) corrrespond to the density computed for silica sodalite (SOD). The correspondence between some of the peaks and three-atom correlations around a framework node is also indicated. 
Note that the correlations included in the SOAP features also incorporate some ``self-correlation'' 
for the density centered on each atom, which have been eliminated here for clarity of visualization.}
    \label{fig:soap-sodalite}
\end{figure*}

This kind of correlation function is often used in the statistical mechanical treatment of liquids, and is typically computed as an average over an ensemble of configurations. In the case of SOAP, however, each individual atomic environment $\CX_j$ is associated with a feature vector $\bra{\alpha r \alpha' r'\omega}\ket{\CX_j}$ that provides a unique~\cite{bart+13prb} description of a given arrangement of atoms. For example, Fig.~\ref{fig:soap-sodalite} shows a plot of $\bra{\ce{Si} r \ce{Si} r'\omega}\ket{\CX_j}$ and  $\bra{\ce{O} r \ce{O} r'\omega}\ket{\CX_j}$ for a Si atom in the framework of sodalite, indicating the correspondence between peaks in the correlation function and Si-Si-Si and O-Si-O atomic motifs.

The representation $\bra{\alpha r \alpha' r'\omega}\ket{\CX_j}$ can be expanded on an orthogonal basis built out of radial functions $R_n(r)$ and Legendre polynomials $P_l(\omega)$ ~\cite{will+19jcp}, leading to the usual form of SOAP~\cite{bart+13prb} as a power spectrum $\bra{\alpha n \alpha' n' l}\ket{\CX_j}$, which is the projection of the three-body density onto radial basis functions $n$ and $n'$ and Legendre polynomial $l$. The numbers of radial and angular functions are convergence parameters, and the two parameters that are more significant in determining the behavior of the SOAP representation are the cutoff distance of the region included in the environment and the width $\sigma$ of the Gaussians, which controls how a difference in atomic coordinates between two structures translates in a difference between their respective $\bra{\alpha r \alpha' r'\omega}\ket{\CX_j}$ vectors. Below we consider cutoff distances of 3.5 \AA{} and 6.0 \AA{}, leading to SOAP vectors with approximately 3,000 elements. In the Supporting Information, we list the remaining parameters needed to specify our application of SOAP.

\subsection{\label{sec:MLmodel} Machine learning model and environment-centered properties}

To assess the performance of the different representations in quantifying structure--property relations, we built a machine learning model to predict the molar volume and the cohesive energy of each structure based on the sum of environment feature vectors defined by each representation. The performance of the model, and the rate at which the error on a test dataset decreases as the model is trained on increasing amounts of data (so-called learning curves), offer insights into the completeness and directness of a representation for describing a particular property or set of properties~\cite{Huang2016,vonLilienfeld2015,Faber2017}. 
For the property prediction we use a statistical learning scheme based on Kernel Ridge Regression (KRR), a relatively simple approach that makes it possible to introduce some non-linearity in the regression while having a deterministic learning procedure. Kernel Ridge Regression has been used successfully to build models for several classes of atomic-scale properties~\cite{Schutt2014,Rupp2012,Hansen2013,Hansen2015,Faber2015,Faber2016,Faber2017,vonLilienfeld2015}, and a formally equivalent method, Gaussian Process Regression, has been used to create interatomic potentials~\cite{Bartok2010,Bartok2015,Bartok2018,Szlachta2014,Dragoni2018,Maillet2018,Deringer2017}.

For each descriptor, we built machine learning models for the molar volume and energy, using a reference data set that we introduce in Section \ref{sec:data}. To obtain a model that can be evaluated at fixed cost and thus independent on the training set size, we used a sparse KRR model~\cite{Smola2000,rasm06book,Ceriotti2018} employing a Gaussian kernel and based on a set of 2,000 representative environments $M=\{\CX_j\}$ selected with Farthest Point Sampling (FPS)~\cite{elda+97ieee,ceri+13jctc,Imbalzano2018}.
\footnote{For some of the ring descriptors, the dataset contains fewer than 2,000 unique feature vectors, in which case  only the unique representations were considered as representative environments.}
For each regression task we optimized the parameters of the kernel and the regression model through a grid search and five-fold cross validation with the aim of minimizing the mean absolute error (MAE) of the regression. Additional details of our approach are discussed together with each example in the Supporting Information.

The sparse KRR model we use approximates the target property $y$ for a zeolite framework $\CA$ as a sum over contributions from each environment $\CX$ in that framework, i.e.,
\begin{equation}
y(\CA)=\sum_{\CX\in\CA} y(\CX) = 
\sum_{\CX\in\CA} \sum_{\CX_j\in M} x_j k(\CX,\CX_j),
\end{equation}
where $x_j$ is a weighting factor and $k(\CX, \CX_j)$ is the kernel between the local environments $\CX$ and $\CX_j$.
As a consequence, the regression exercise also permits decomposition of structure-wide properties (like volume or energy) into contributions from the individual atomic environments $y(\CX)$, which can be used to investigate and visualize the impact of local structural motifs on macroscopic properties of the framework.

\subsection{\label{sec:dimred} Dimensionality reduction}
While machine learning models based on kernel methods can be quite powerful, they can also be computationally expensive, especially for very large datasets. When working with large amounts of data, as we do here, it is often necessary to sparsify the kernel used in the machine learning model. To this end we use a low rank approximation to the true kernel matrix built by considering the kernel between each environment and a set of representative environments. 
We selected the representative environments with an iterative strategy based on FPS~\cite{elda+97ieee,ceri+13jctc,Imbalzano2018},
as noted in Section \ref{sec:MLmodel}, to ensure that the representative environments are structurally diverse. We used approximate kernel matrices constructed in this way to perform KRR and kernel principal component analysis (KPCA) \cite{Scholkopf1998}. We applied KPCA to the SOAP-based descriptors to reduce the dimensionality of the descriptor and to assess the information content of the descriptor in the context of predicting the energy and volume of hypothetical zeolite structures.
Additional details regarding the kernel approximations for KRR and KPCA are given in the Supporting Information.

\section{\label{sec:data} Data Selection}
Our analysis of zeolite structures is based on the DEEM SLC-PCOD database \cite{Pophale2011},
which contains hypothetical zeolite structures that are no more than 30 kJ/mol Si higher in energy than $\alpha$-quartz. 
Given that the database contains a few hundred thousand hypothetical zeolites, each zeolite contains several \ce{Si}-centered environments, and that a single SOAP feature vector can contain thousands of components, computing and analyzing the full SOAP vectors of every structure in the database is computationally intractable at this time. Hence, we reduced the dimensionality of the input space by considering a subset of the database and by selecting only the most diverse SOAP components to use in our ML model, as explained in Section \ref{sec:dimred}.
From the approximately 330,000
structures in the database, a subset of 10,000 structures was selected at a fixed stride following the ID number, and
the 500 most diverse SOAP vector components were selected via FPS on a random selection of 2,000 structures from the 10,000-structure subset. The Euclidean distance between SOAP vectors was used as the distance metric for the FPS procedure \cite{Imbalzano2018}. The SOAP vectors were then computed for all 10,000 structures in the subset, but only the FPS components were retained.
Through this scheme, we considered two SOAP descriptors: one with a radial cutoff of 3.5 \AA{}, and another with a 6.0 \AA{} cutoff, so that the representation incorporates information on the size scale of nearest neighbor \ce{Si} atoms (3.5 \AA{}) or next-nearest neighbors, possibly including a modestly sized ring (6.0 \AA{}). Each atomic environment comprised a central \ce{Si} atom and all of the surrounding \ce{Si} and \ce{O} atoms within the cutoff radius. Oxygen atoms were not considered as environment centers. Additional details regarding the construction of the SOAP vectors are given in the Supporting Information.
The distance-, angle-, and ring-based descriptors were also calculated for the 10,000-structure subset for comparison against SOAP in our machine learning models. Comparisons between the classical and SOAP descriptors are examined in the following section.

In addition, we considered a subset of 1,000 stride-selected structures from the DEEM SLC-PCOD database for our analysis to test whether our results are influenced by the size of our selected subset of structures. We found that the results for the 1,000-structure sample yield similar conclusions to those that can be drawn for the 10,000-structure sample, both in terms of performance when being used as the basis of a machine learning model to predict different geometric features, and in terms of the representation of the data set using unsupervised learning. This robustness with respect to dataset size indicates that our results have likely converged with respect to the structural diversity in the Deem database, and thus, that our results apply to the full Deem database.
In Section \ref{sec:results} we report our findings based on the 10,000-structure subset. The results for the 1,000-structure subset can be found in the Supporting Information.

\section{\label{sec:results} Results }

\subsection{\label{sec:descriptor} Comparison of Structural Representations}

Many machine learning studies present parity plots to show the quality of learning for a given property for the best model. We prefer to quantify learning by calculating mean average errors (MAEs) of a property estimation, and presenting MAEs as learning curves, i.e., as functions of the number of training points.
Building learning curves for the prediction of molar volume and energy enables head-to-head comparisons of the information content in the various structural descriptors. In particular, the value of the MAE for small training set sizes indicates whether the most prominent components of a representation correlate strongly with a given property. The asymptotic behavior in the large training set sizes indicates how complete a representation is: saturation of a learning curve, i.e., a learning curve that flattens to nearly zero slope, indicates that there is not sufficient information to improve the model, even when exposing it to new data; while a substantial asymptotic slope indicates that the model still has sufficient information to improve its learning.

\begin{figure}
\centering
\includegraphics[width=0.45\textwidth]{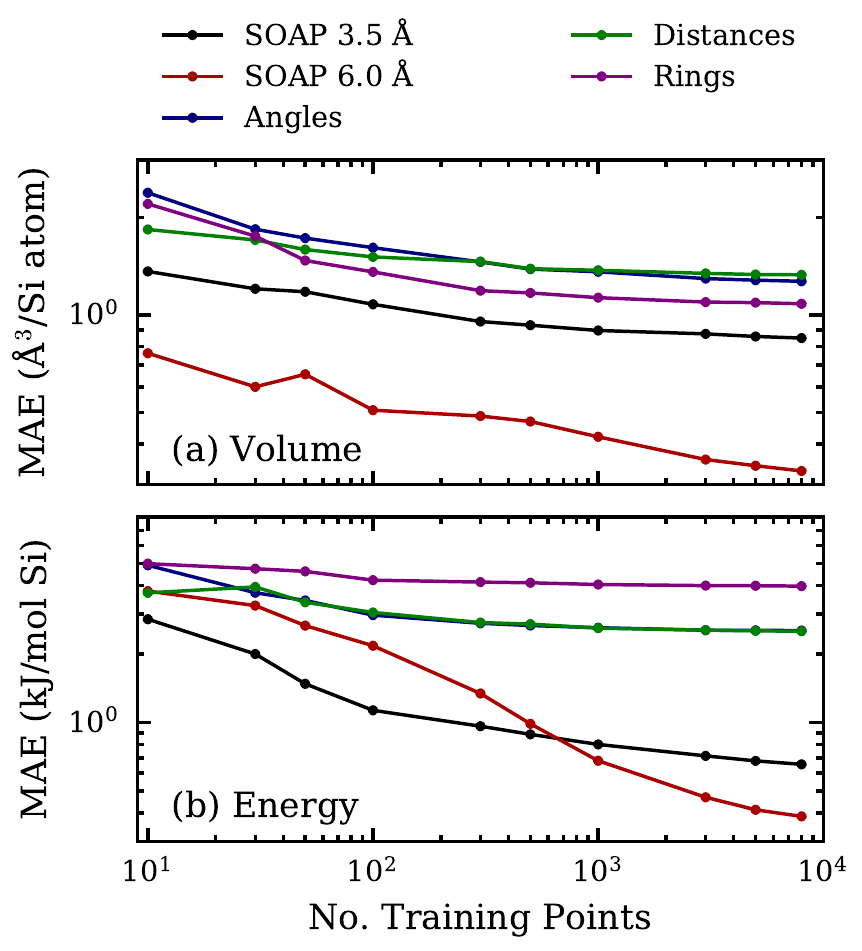}
\caption{Learning curves of the classical and SOAP descriptors for predictions of (a) volume per Si atom and (b) energy per mol \ce{Si}.}
\label{fig:learn}
\end{figure}

Figures \ref{fig:learn}(a) and \ref{fig:learn}(b) show the learning curves for predictions of the molar volume and molar energy, respectively, using the classical and SOAP descriptors. We see in Figures \ref{fig:learn}(a) and \ref{fig:learn}(b) that the \ce{Si-Si} distance (green), 
\ce{Si-O-Si} angle (blue), and shortest-path rings (violet) perform poorly at predicting both volume and energy, saturating at MAEs above 1 \AA$^{3}$/Si atom and 2 kJ/mol Si, respectively.
The ring-based descriptor is marginally better at predicting volumes compared to the distance and angle descriptors [see \ref{fig:learn}(a)], but it is the worst classical descriptor at predicting energies [see \ref{fig:learn}(b)]. 
This is likely because lattice energy is sensitive to ring geometries and distortions, thus requiring more information than simply the numbers of atoms in rings.

Using the SOAP descriptor with a 6.0 \AA{} cutoff (red) results in the best volume predictions for all training set sizes, with MAEs as low as 0.4 \AA$^{3}$/\ce{Si} atom.
This is probably due to the fact that accurate predictions of overall framework densities require information on larger spatial scales. The SOAP descriptor with a 3.5 \AA{} cutoff (black) performs only slightly better than the classical descriptors in predicting volumes, but yields the best energy predictions off all the descriptors for smaller training set sizes, indicating that relatively local correlations are sufficient for making estimates of lattice energy within 1 kJ/mol Si. This is perhaps not surprising since a substantial contribution to the zeolite lattice energy can be accounted for through nearest-neighbor bond and angle interactions, and thus the 3.5 \AA{} based model, where only the first neighboring tetrahedral information is included, does a good job at estimating the energy even when the training set has a modest size. However, the learning potential of the 3.5 \AA{} SOAP descriptor plateaus as the training set size approaches 1,000 structures, while the 6.0 \AA{} SOAP representation continues to improve at larger training set sizes, yielding energy predictions with MAEs as low as 0.4 kJ/mol Si. Overall, Figures \ref{fig:learn}(a) and \ref{fig:learn}(b) show that the SOAP descriptor with a 6.0 \AA{} cutoff does an excellent job of accurately capturing zeolite energy and molar volume, while the classical descriptors do not.

\begin{figure}
    \centering
    \includegraphics[width=0.45\textwidth]{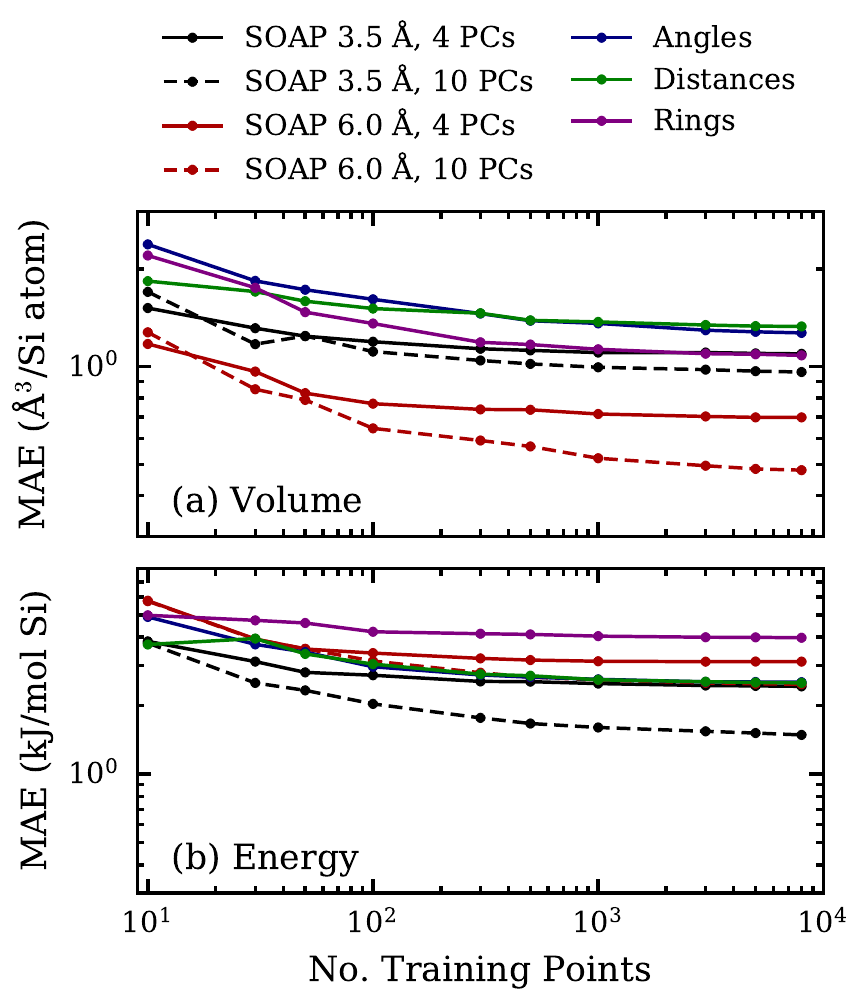}
    \caption{Learning curves of classical and SOAP-KPCA descriptors with similar dimensionality for predictions of (a) volume and (b) energy.
    }
    \label{fig:learnKPCADim10k}
\end{figure}

To investigate whether the improved predictive performance and learning ability of the 6.0 \AA{} SOAP descriptor is a result of the intrinsic quality of the descriptor itself, or merely from the fact that the 6.0 \AA{} SOAP representation incorporates more information through higher-dimensional vectors, we compare the learning curves of the classical descriptors with those of the SOAP descriptors whose dimensionality has been reduced through kernel principal component analysis (KPCA). Because the distance- and angle-based descriptors are represented as 4-element vectors (because of the tetrahedral coordination around \ce{Si}) and the ring vectors contain ten elements (because ring sizes in our data set range between 3-rings and 12-rings), we find it instructive to compare the classical descriptors with the first four and ten principal components of the SOAP representations, shown in Figs.\ \ref{fig:learnKPCADim10k}(a) and \ref{fig:learnKPCADim10k}(b). For the volume prediction (see \ref{fig:learnKPCADim10k}(a)), the SOAP-KPCA 6.0 \AA{} including ten dimensions (red, dashed) is the best model, followed by the SOAP-KPCA 6.0 \AA{} including four dimensions (red, solid), which has comparable performance to the ten-dimensional representation for training sets containing fewer than 50 points (structures). In the case of the energy prediction depicted in Figure \ref{fig:learnKPCADim10k}(b), the best model is that based on the SOAP-KPCA 3.5 \AA{} descriptor including ten dimensions (black, dashed), followed by the four-dimensonal descriptor (black, solid). Overall, we note that SOAP-KPCA models equal or surpass the performance of the classical descriptors at predicting energy and volume, even after dimensionality reduction. 
This suggests that SOAP inherently contains more information about the local structure of a zeolite framework than do the classical descriptors, and this is not a mere result of the flexibility afforded by the higher dimensionality of the feature representation.

For predictions of the molar volume, the SOAP-KPCA descriptors perform better than the classical descriptors regardless of the dimensionality of the representation, though the performance gain of the 3.5 \AA{} SOAP-KPCA descriptor over the classical descriptors is rather small. For predictions of lattice energy per \ce{Si} atom, the 3.5 \AA{} SOAP-KPCA descriptor performs as well as or better than the classical descriptors at larger training set sizes. The 6.0 \AA{} SOAP-KPCA descriptor, on the other hand, performs worse than the distance- and angle-based descriptors at comparable dimensionality, but better than the rings-based descriptor. The behavior of 6.0 \AA{} SOAP-KPCA can be attributed to the fact that 6.0 \AA{} SOAP considers a larger local environment, and therefore, a much larger amount of information about the surroundings of a single \ce{Si} atom. 
This additional information simply becomes noise within the regression model unless more information---in the form of additional principal components---is included.
Indeed, in predictions of the energy per \ce{Si} atom, we find that the 3.5 \AA{} SOAP representation outperforms the 6.0 \AA{} SOAP representation for a given number of principal components unless upwards of 100 components are included. (Learning curves for the SOAP-KPCA representations with a larger range of principal components are provided in the Supporting Information.) 

\subsection{\label{sec:validation} Low-Dimensional Maps of the Deem Dataset}

\begin{figure}
    \centering
    \includegraphics[width=0.45\textwidth]{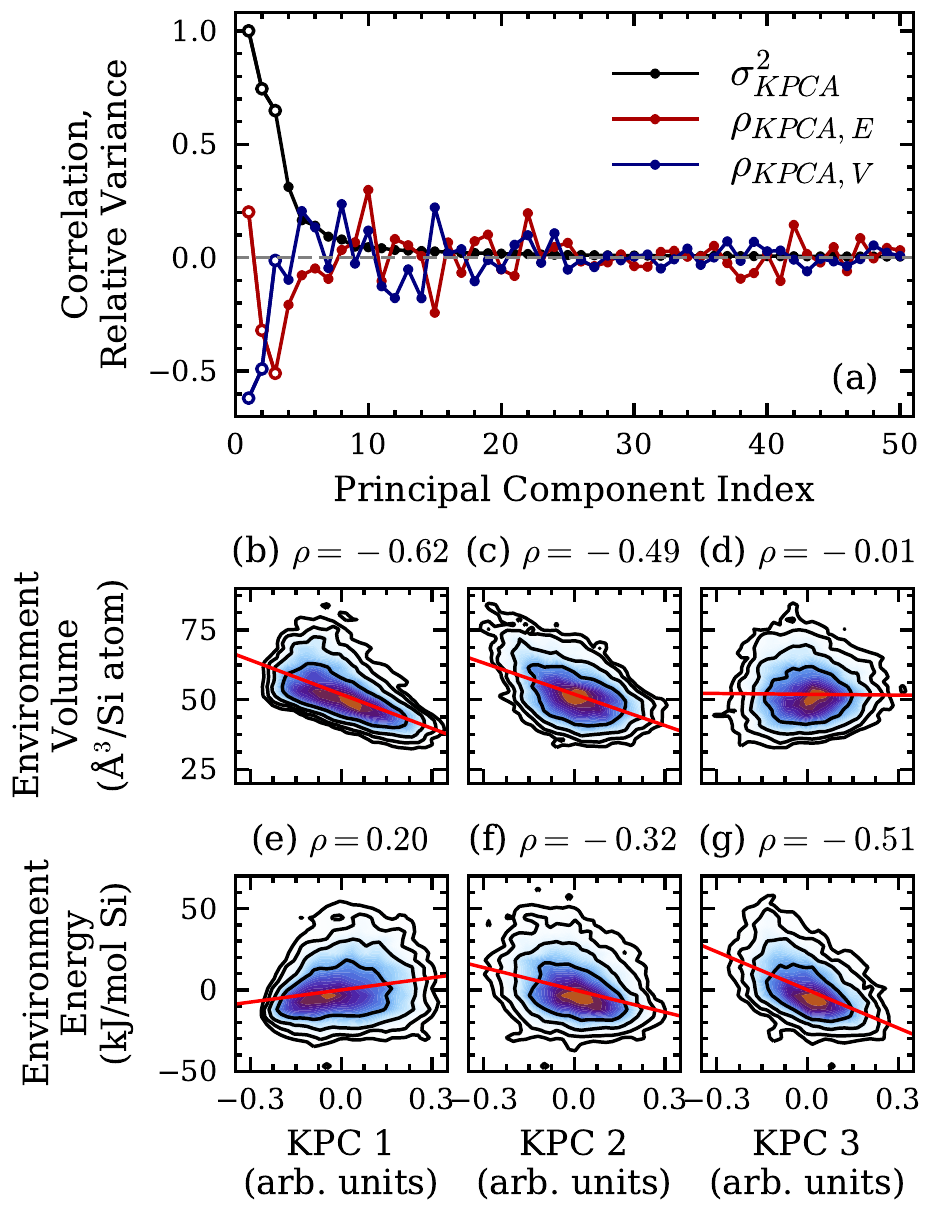}
    \caption{(a) Pearson correlation coefficients between the first 50 KPCs of the 6.0 \AA{} SOAP representation and the decomposed environment volumes and energies in the 10,000-structure sample. The relative variance in the KPCs at each of the first 50 components is also plotted. The correlation coefficients and relative variance of the first three components are highlighted with open symbols, as these are the KPCs that correlate most strongly with the decomposed environment energies and volumes.
    (b)--(g) Kernel density estimation of all environments in KPC--property space with a linear least squares fit to the data to show the correlations in more detail (red line). The value of the Pearson correlation coefficient is given above each contour set.}
    \label{fig:correlation}
\end{figure}

The results in Fig.\ \ref{fig:learnKPCADim10k} show that the SOAP method provides sufficient information, even when truncated in dimensionality, to outperform classical descriptors in {\it {quantitatively}} accounting for the molar volume and energy of a database of zeolites. This finding raises the following question: how many dimensions of SOAP are required to {\it {qualitatively}} account for the structural diversity found in the Deem database? To answer this question, we show in Fig.~\ref{fig:correlation}(a) the relative variances ($\sigma^2_{KPCA}$) of the kernel principal components (KPCs) and their Pearson correlation coefficients with energy ($\rho_{KPCA,E}$) and volume ($\rho_{KPCA,V}$). The KPCs are sorted, by construction, in a way that reflects the intrinsic variability of the dataset, and the first three components in Fig.\ \ref{fig:correlation}(a) (open circles) capture a substantial portion of the structural diversity as evidenced by the relative variances all exceeding 0.6, while all other relative variances are less than 0.3 and most are much less than 0.1. The Pearson correlation coefficients in Fig.\ \ref{fig:correlation}(a) reveal that KPCs 1 and 2 correlate strongly with volume while KPCs 2 and 3 correlate strongly with energy.  These correlations can also be visualized in the contour plots shown in Fig.~\ref{fig:correlation}(b)--(g), where the environment energies and volumes are plotted against values of the first three KPCs along with a linear least squares fit to the data.
The elongated shapes of the contours indicate the correlations between the environment properties and the SOAP vectors in KPC space, and these correlations are quantified by the Pearson correlation coefficients.
Overall, Figs.~\ref{fig:correlation}(a)--(g) suggest that a three-dimensional picture of structural diversity using the first three KPCs can reveal the essential features of the Deem zeolite data set.

Consequently, we can use the SOAP-KPCA construction to build a new 3D ``atlas'' of zeolite building blocks, analogous to the list of 60 composite building units (CBUs) in the IZA database \cite{ZeoAtlas2007,ZeoAtlasCBUs}, and to the list of ``packing units'' given by Blatov {\it {et al}}. \cite{Blatov2013}. As emphasized above, virtually all CBUs and packing units are centered on void spaces; in contrast, our zeolite building blocks are atom-centered, allowing such environments to be summed to yield macroscopic properties of overall frameworks such as molar energy and volume. Because of the very large number of local environments in the Deem database, our new atlas has the feel of a point cloud in 3D-KPCA space. 

\begin{figure*}
    \centering
    \includegraphics[width=\textwidth]{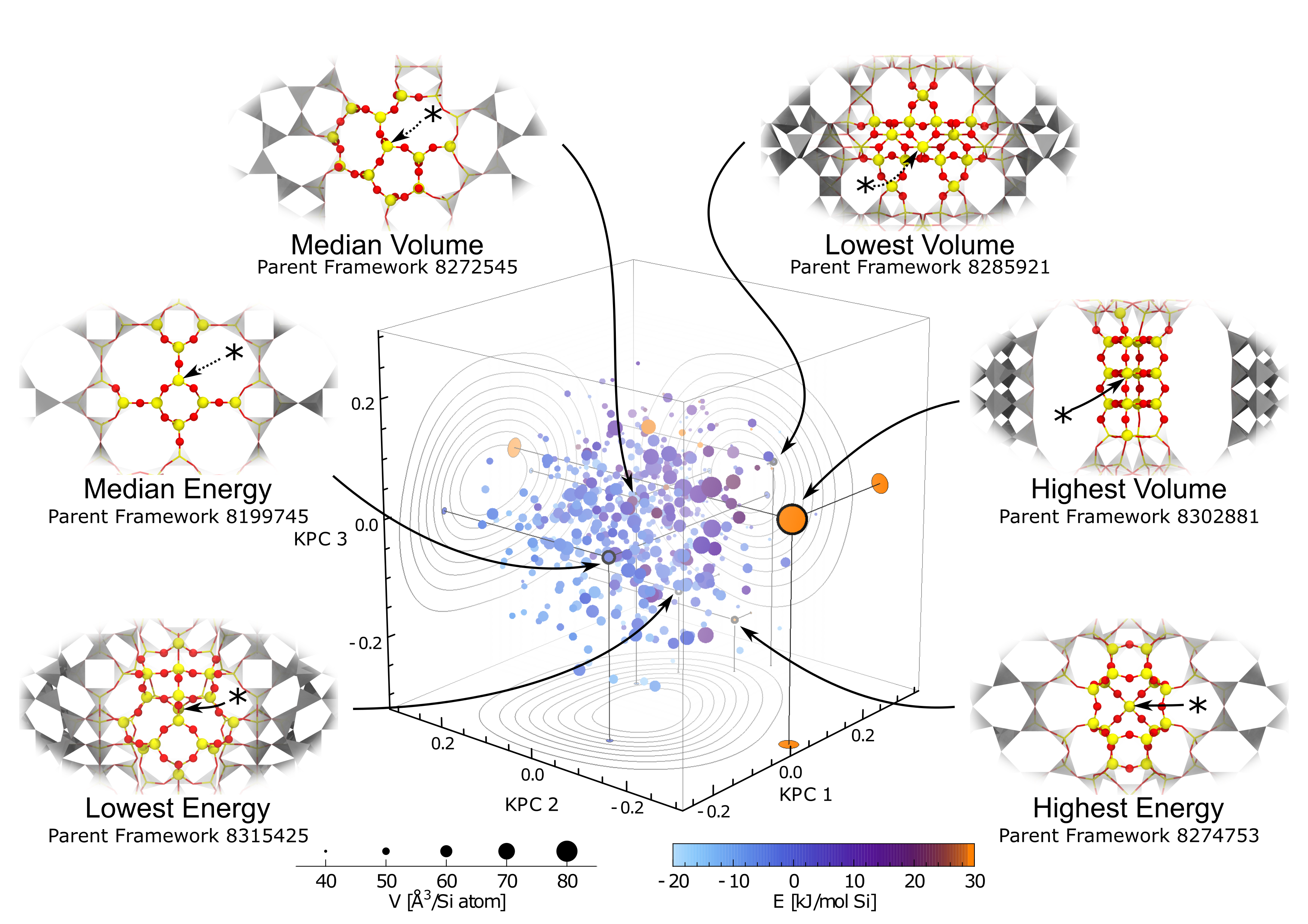}
    \caption{The new atlas of zeolite building blocks, where every 2,000-th environment of the 10,000-framework subset is plotted as a point in the three-dimensional space formed by the first three kernel principal components of the SOAP construction using a 6.0 \AA{} cutoff. The points are colored and sized according to the energy and volume contribution of the corresponding environment. The environments with the highest and lowest energies are highlighted along with environments contributing energies and volumes close to the median of the dataset. Note that there exist some (extreme) outliers: the highest-energy environment contributes more than 380 kJ/mol \ce{Si}, and the lowest below --30 kJ/mol \ce{Si}. The highest-volume environment contributes more than 90 \AA{}$^3$/Si atom, and the lowest less than 30 \AA{}$^3$/Si atom. Energies falling outside the range of the scalebar are assigned to the color at the nearest extreme of the colorscale. Environment centers are indicated by the asterisks and their associated arrows; a dotted arrow signifies that the central atom is hidden behind the foremost atom visible in the atomic snapshot. In each snapshot, the atomic environment is represented as a ball-and-stick model; the surrounding zeolite structure is represented as \ce{SiO2} tetrahedra. Overall, we see a remarkably uniform distribution of environments.}
    \label{fig:atlas}
\end{figure*}

Our SOAP-based ``cloud atlas'' is presented in Fig.\ \ref{fig:atlas}, where the SOAP environments are plotted in the space defined by the first three KPCs of the dataset. Each point (environment) is colored according to its energy contribution (low = white, medium = blue/purple, high = orange) and is sized according to the volume contributions to its corresponding zeolite framework. Several environments are highlighted to provide examples of the ``building blocks'' present in this new cloud atlas. In particular, we show in Fig.\ \ref{fig:atlas} the lowest-~, median-~, and highest-energy SOAP environments, as well as the lowest-, median-, and highest-volume SOAP environments.
The majority of environment volumes in Fig.\ \ref{fig:atlas} range from 30--80 \AA$^3$/Si, corresponding to a range of framework densities of 12.5--33.3 Si atoms per 1000 \AA$^3$. (Actual zeolites exhibit framework densities in the range of 12.5--21.0 Si atoms per 1000 \AA$^3$ \cite{ZeoAtlas2007}, although the Deem database extends to larger framework densities than those of actual zeolites.) Most environment energies fall within the range of --20 kJ/mol Si to +30 kJ/mol Si.
We note that the notions of environment energy and volume relate herein to the machine learning process and not necessarily to physico-chemical notions of energy or volume. 
Nevertheless, as we will show later, this data-driven decomposition ends up being consistent with physical considerations, as was already observed in similar ML models of materials and molecular properties~\cite{Bartok2017,wilk+19pnas}.
In each highlighted environment, the atoms included within 6.0 \AA{} SOAP are shown in yellow (\ce{Si}) and red (\ce{O}), while the rest of the zeolite framework is depicted as corner-sharing tetrahedra. Contour plots showing the distribution of environments are projected onto the $xy$-, $yz$-, and $xz$-planes in Fig.\ \ref{fig:atlas}. These contour plots reveal that the statistical distribution of zeolite environments is unimodal and rather broadly peaked, indicating a remarkably uniform distribution of structural motifs. As such, we find no special region in 3D KPC-space that is particularly well stocked with building blocks for making hypothetical zeolites. 
The broad distribution of environments shown in Fig.\ \ref{fig:atlas} suggests that the algorithm for producing these hypothetical zeolite frameworks \cite{Deem2009,Pophale2011} has left no substantial gap in environment space.

\begin{figure}
    \centering
    \includegraphics[width=0.45\textwidth]{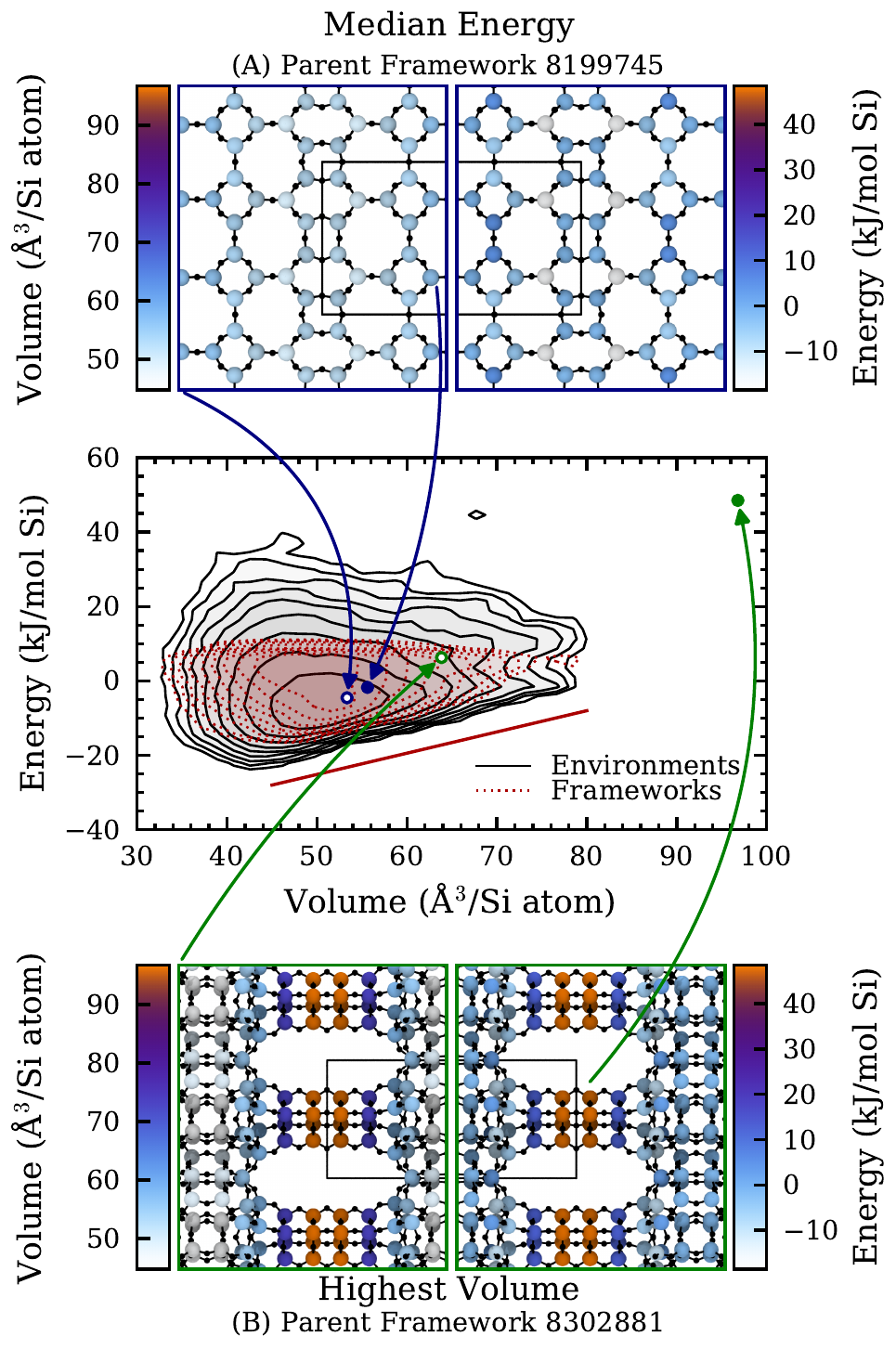}
    \caption{
    Middle panel: contour plots showing probabilities of finding frameworks (red dotted lines) and environments (solid black lines) with certain values of molar energy and volume in the Deem dataset, using the SOAP 6.0 \AA{} representation on the 10,000-structure sample for the environment analysis. Comparing framework and environment contours shows similar structure below 10 kJ/mol Si, with the same energy--volume correlation highlighted by a thick red line.
    Top and bottom panels: hypothetical frameworks labeled A and B, with median-energy (in framework A) and highest-volume (in framework B) environments, corresponding to solid blue and green dots, respectively, in the contour plot. Parent frameworks A and B are labeled by open blue and green dots, respectively, in the contour plot.
    Left and right panels show local volume and energy distributions, respectively, for each framework. Visual inspection shows that framework B exhibits much more heterogeneity in both local energy and volume, with strong local energy--volume correlations.
    }
    \label{fig:EnvFrame_6.0_10k}
\end{figure}

A complementary way to visualize information contained in the Deem dataset is by constructing energy--volume contour plots for both frameworks and their constituent environments; this is shown in the middle panel of Fig.~\ref{fig:EnvFrame_6.0_10k}, with environment contours in solid black and framework contours in dotted red. Each line in these contour plots indicates a region of constant probability for finding a framework, or an environment, with a certain combination of molar energy and volume. These contour plots show an excellent overlap between framework and environment spaces for energies between --20 kJ/mol Si to +10 kJ/mol Si. Indeed, Fig.~\ref{fig:EnvFrame_6.0_10k} shows that most local volumes obtained by machine learning map very well onto the actual space of framework volumes in the Deem dataset. Furthermore, the lower edges of both contour plots (highlighted by the thick red line) follow the energy-density correlation known for actual zeolites \cite{Henson1994}. This finding has already been noted by Deem and coworkers \cite{Pophale2011} for the hypothetical frameworks; we find it remarkable that the SOAP-based environments also exhibit the signature of this energy--volume correlation.

The contour plots in Fig.~\ref{fig:EnvFrame_6.0_10k} also show a significant space of environments with local energies greater than 10 kJ/mol Si, i.e., greater than that of any framework in our sample of the Deem database, which was constructed to cover an energy range no more than 30 kJ/mol Si above the molar energy of $\alpha$-quartz \cite{Pophale2011}. This extension to higher local energies raises the question of how high-energy local environments manifest in overall framework structures. Such a question on the connection between local environments and overall frameworks can also be raised for high-volume environments, which presumably frame large pores. The top and bottom panels in Fig.~\ref{fig:EnvFrame_6.0_10k} answer these questions for two hypothetical frameworks labeled A (top) and B (bottom). Each panel has two copies of each framework showing local volumes (left) and local energies (right), both color-coded according to low = white, medium = blue, and high = orange. 
Framework A (top) is shown because it contains the median-energy environment in the Deem dataset, corresponding to the solid blue dot in the contour plot, while its parent framework (A) corresponds to the open blue dot.
Framework B (bottom) was chosen because it contains the highest-volume environment in the Deem dataset, corresponding to the solid green dot in the contour plot, while its parent framework (B) corresponds to the open green dot.

These energy-volume color-coded images provide a qualitatively new way to visualize zeolite frameworks, yielding several interesting insights.
First, by comparing the color distributions in frameworks A and B, it is readily apparent that A (a medium-pore framework) is much more homogeneous in both local energy and volume than is B (an ultra-large-pore framework). 
Second, we see that the indicated environment in B (centered on the orange-colored Si atom) does indeed frame the ultra-large pore. Surprisingly, though, the other Si atoms that frame this same ultra-large pore exhibit lower local volumes, presumably because these other Si atoms anchor environments closer to denser regions of the framework and its pore walls.
Third, a visual comparison of the volume (left) and energy (right) representations of framework B shows clear correlations between local volumes and energies, reminiscent of the correlation for actual zeolites reported by Henson and coworkers \cite{Henson1994}. Such a correlation is harder to discern for framework A because of its homogeneous character. 
Overall, these color-coded images, especially those for framework B, show that local energies and volumes vary smoothly across a given framework, possibly allowing the visualization of framework strain and other materials and chemical properties, to be studied in a forthcoming publication.

\section{\label{sec:conclusions} Conclusions}

We have presented an analysis of the structural motifs found within a database of hypothetical zeolites, the Deem SLC-PCOD database \cite{Pophale2011}. We have investigated whether the structural diversity found in the Deem database can be well-represented by classical descriptors such as Si-Si distances, Si-O-Si angles, and ring sizes, or whether a more general-purpose geometric representation of atomic structure, herein afforded by the SOAP method, is required. We assessed the quality of each descriptor by determining its ability to inform the machine-learning of molar energy and volume for each hypothetical framework in the dataset. We have found that a SOAP representation with a cutoff-length of 6 \AA{}, which captures local environments beyond near-neighbor tetrahedra, best describes the structural diversity in the Deem database as measured by the accurate predictions of energy and volume afforded by this approach. 
The performance gain of SOAP can be traced mainly to its description of all inter-atomic correlations within an atom-centered environment. The SOAP approach was found to maintain its lead even when reducing the dimensionality of SOAP feature vectors, through kernel principal component analysis, to be comparable with those of Si-Si distances, Si-O-Si angles, and ring sizes. We thus find that SOAP provides a powerful tool for the statistical analysis of local zeolite environments, i.e., the building blocks of zeolites.

SOAP-KPCA also showed that the first three kernel principal components capture the main variability in the data set, allowing a 3D point cloud visualization of local environments in the Deem database. We refer to this visualization as a ``cloud atlas'' of zeolite building blocks.
This cloud atlas was found to show
good correlations with the contribution of a given motif to the density and stability of its parent framework. 
Overall, the cloud atlas of the Deem database shows a uniform distribution of structural motifs, suggesting that the algorithm for producing the Deem database covered structure space in a thorough and egalitarian fashion.
Local volume and energy maps constructed from the SOAP/machine-learning analyses provide new ways of visualizing zeolites that reveal structural heterogeneity, smooth variations of local volumes and energies across a given framework, and correlations between volume and energy of local environments within a given framework.

The present work raises many questions in both data science and zeolite science. From a data science perspective, we wonder whether implementing SOAP with multiple length scales~\cite{bart+17sa} or with radial scaling~\cite{will+18pccp} could optimize the packaging of zeolite structural information, perhaps at the expense of introducing further optimization parameters and obscuring some of the insights discovered above. From a zeolite science perspective, several questions arise, including whether the framework density/stability maps introduced above can predict important zeolite properties like acid-site strength across a given framework. It is also interesting to consider applying the SOAP/machine-learning methods to actual zeolites in the IZA database to determine similarities and differences between actual- and hypothetical-zeolite data sets. Finally, it is interesting to consider applying the machinery of SOAP to simulations of zeolite formation \cite{Auerbach2015,Bores2018} to determine the local environments that eventually lead to zeolite crysallization. These topics will be addressed in forthcoming contributions.

\begin{acknowledgments}
BAH and MC were supported by the European Research Council
under the European Union's Horizon 2020 research and innovation programme
(Grant Agreement No. 677013-HBMAP). RS acknowledges support by the SCCER Efficiency of Industrial Processes. G.P. is thankful to Regione Autonoma della Sardegna for the Ph.D. fellowship provided under the “POR-F.S.E. 2014-2020” program.
All the Authors would like to acknowledge The National Centre of Competence in Research (NCCR) \lq\lq Materials' Revolution: Computational Design and Discovery of Novel Materials (MARVEL)\rq\rq ~of the Swiss National Science Foundation (SNSF), which supported a visit of SMA to EPFL, during which this collaboration was initiated.

\end{acknowledgments}

\clearpage
\ \vfill
{\centering \Huge Methods\\}
\vfill
\foreach \x in {2,3,4,5,6,7,8,9,10,11,12}
{%
\clearpage
\includepdf[pages={\x}]{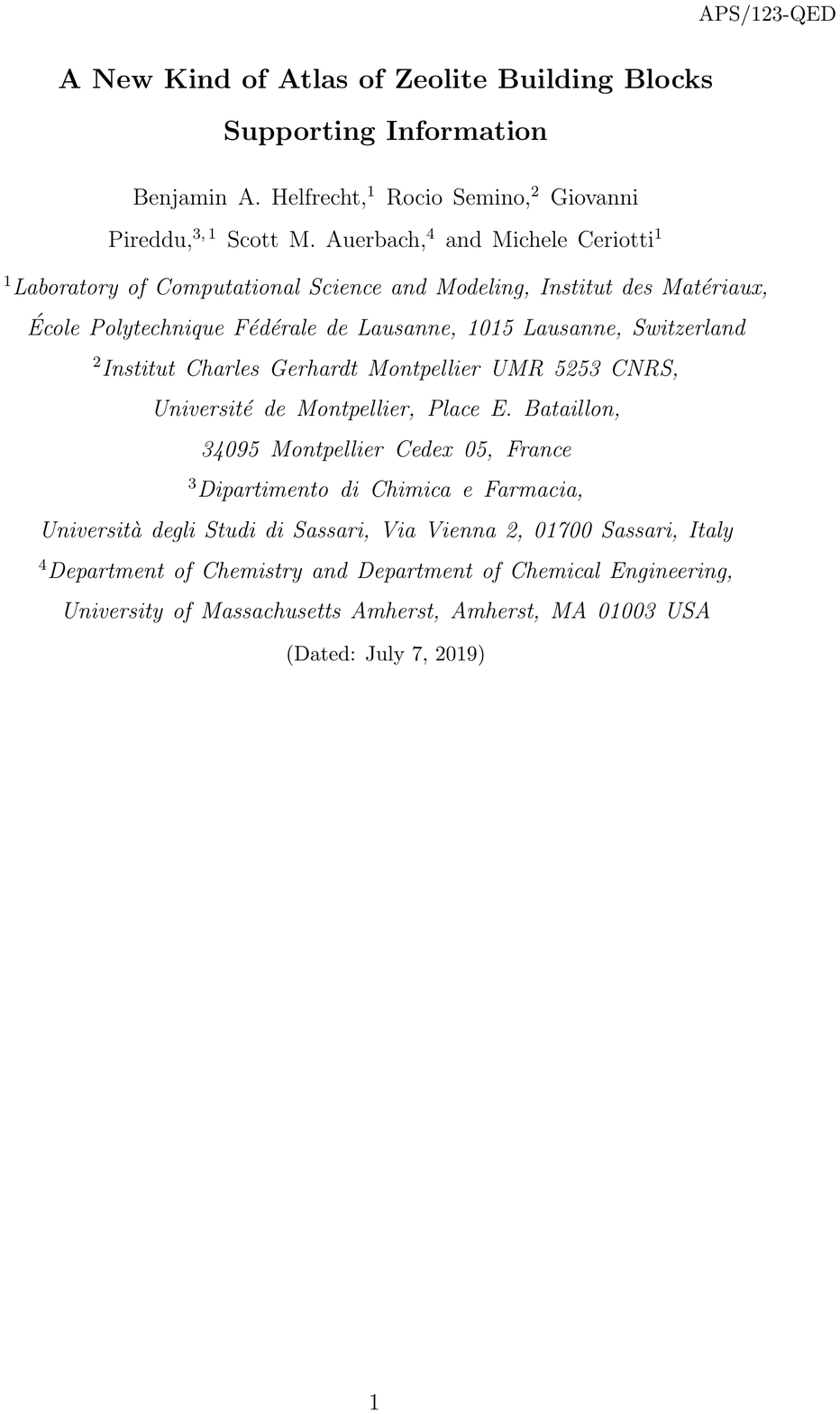}
}

\end{document}